\documentclass
[aps,pra,twocolumn,twoside,amssymb,footinbib,groupedaddress]{revtex4}%
\usepackage{amsmath}
\usepackage{graphicx}%
\usepackage{amsfonts}%
\usepackage{amssymb}
\newtheorem{theorem}{Theorem}

\newtheorem{definition}[theorem]{Definition}

\newtheorem{lemma}[theorem]{Lemma}

\newtheorem{Result}{Result}

\newenvironment{proof}[1][Proof]{\textbf{#1.} }{\ \rule{0.5em}{0.5em}}

\begin{document}

\title{Characterization of non-local gates}
\author{K. Hammerer$^{1}$, G. Vidal$^{2}$ and J.I. Cirac$^{1}$}
\affiliation{$^{1}\,$Max-Planck Institut f\"{u}r Quantenoptik, Hans-Kopfermann Str. 1,
D-85748 Garching, Germany.}
\affiliation{$^{2}\,$Institute for Quantum Information, California Institute of Technology,
Pasadena, CA 91125, USA.}

\begin{abstract}
A non-local unitary transformation of two qubits occurs when some Hamiltonian
interaction couples them. Here we characterize the amount, as measured by
time, of interaction required to perform two--qubit gates, when also
arbitrarily fast, local unitary transformations can be applied on each qubit.
The minimal required time of interaction, or \emph{interaction cost}, defines
an operational notion of the degree of non--locality of gates. We characterize
a partial order structure based on this notion. We also investigate the
interaction cost of several communication tasks, and determine which gates are
able to accomplish them. This classifies two--qubit gates into four
categories, differing in their capability to transmit classical, as well as
quantum, bits of information.

\end{abstract}
\volumeyear{year}
\volumenumber{number}
\issuenumber{number}
\eid{identifier}
\date{\today                     }
\startpage{1}
\endpage{2}
\maketitle

\section{Introduction}

An essential ingredient
in quantum information processing is the ability to make two two--level
systems or qubits undergo a joint unitary evolution. Accordingly, most current
proposals for the implementation of a quantum computer rely on some ingenious
method to realize two--qubit gates.

Irrespective of the physical substrate of the qubits, a joint unitary
evolution can only be achieved through some form of interaction. This quite
often couples the two qubits directly, though a third system may alternatively
mediate in the transformation. The starting goal of this paper is, given any
fixed two--qubit Hamiltonian, to describe how it can be used to accomplish any
desired gate on the two systems.

Of course, some form of external control on the two qubits is required to
conveniently modify their evolution, which would otherwise be dictated only by
the coupling interaction. Inspired by the possibilities presently demonstrated
in several quantum optical setups, where each qubit can be independently
addressed \cite{qic}, we assume here the ability to perform arbitrary local
unitary operations (LU) on each of the systems. More specifically, we shall
analyze the fast control limit, in which these control operations can be
performed \emph{instantaneously}. Physically, such a limit amounts to assuming
a neat separation between the time scale of the interaction (which is
comparatively slow) and that of the external manipulations.

The setting we consider corresponds, thus, to the so--called gate simulation
under LU of \cite{ben}. This setting has been previously considered in Ref.
\cite{kha}, where powerful mathematical techniques were developed to study
time--optimal strategies; that is, strategies that perform the desired gate by
using the available interaction for the shortest time. In Ref. \cite{vid}, and
by elaborating on the results of \cite{kha} and of \cite{ben,kra,due},
time--optimal strategies have been analytically characterized for any
interaction and gate of two qubits.

The main result of \cite{vid} permits therefore to assess explicitly the
minimum time an interaction is required to simulate a given gate, a measure
that has been called the \emph{interaction cost} of the gate. The merit of
such a measure is twofold: On the one hand, time is by itself a crucial
parameter in present experiments. In order to successfully process quantum
information, unitary evolutions must in practice be enforced in a sufficiently
small time as compared to the decoherence time of the quantum systems. In
several settings, the time--scale of gates is essentially determined by the
interaction between qubits, for one--qubit unitary transformations can be
performed much faster. Then, an efficient use of the interaction becomes a
priority. On the other hand, the minimal realization time or interaction cost
of a gate can be naturally used to compare gates, thereby endowing the set of
non--local transformations with a partial order structure that refers to the
amount of inherent interaction. This, in turn, provides us with a meaningful
notion of the degree of non-locality of a gate, built upon the observation
that local gates can be performed without any interaction.

In the present paper we first reproduce and extend the results of \cite{vid}
concerning the time optimal use of interactions, and put these into work by
characterizing the information exchange associated to a two--qubit gate. In
\cite{vid}, the derivation of the interaction cost rested on a previous proof
of \cite{kha} which requires familiarity with several facts of differential
geometry. Here we present an alternative, self--contained proof, which in
addition employs ideas and a formalism that we believe to be more common to
quantum information community. This new proof is complemented with an expanded
analysis of the interaction cost of two--qubit gates, including several
relevant examples. The overall result is an operational characterization of
two--qubit gates in terms of the interaction resources needed to perform them.

For any specific information processing task, there may be several gates that
can accomplish it. It is then reasonable to investigate the most efficient way
to accomplish the desired task with a given interaction, that is, to search
for the gate with lowest interaction cost compatible with that task. In
particular, a joint gate can be used to transmit information between the
qubits, and one can study the interaction cost of certain communication tasks,
such as the transmission of classical and quantum bits from one system to the other.

A second main goal of this paper is precisely to characterize the minimal
interaction time required to send classical, as well as quantum, information.
As a by-product, and very much in the spirit of \cite{coll} and \cite{eis},
where information exchange has been used to characterize the non--local
content of certain gates, we obtain a complete classification of two--qubit
gates with respect to their transmission capabilities, thereby supplementing
the original characterization of non-local gates.

The results we present can be summarized as follows:

\begin{itemize}
\item Analytical characterization of the interaction cost of any two--qubit
gate by any two--qubit interaction Hamiltonian, through a new, self--contained
proof (section \ref{sec3}).

\item Analytical characterization, in part of the space of two--qubit gates,
of the partial order structure based on the interaction cost (section
\ref{sec3}).

\item Analytical characterization, for any two--qubit interaction, of the
interaction cost of the following communication processes between two qubits
(section \ref{sec4}):

\begin{enumerate}
\item Transmission of one classical bit: c-bit$_{\text{A}\rightarrow\text{B}}$.

\item Simultaneous, bidirectional transmission of two classical bits:
c-bit$_{\text{A}\rightarrow\text{B}}$ and c-bit$_{\text{B}\rightarrow\text{A}%
}$.

\item Transmission of one quantum bit: q-bit$_{\text{A}\rightarrow\text{B}}$

\item Simultaneous, bidirectional transmission of one classical bit and one
quantum bit: c-bit$_{\text{A}\rightarrow\text{B}}$ and q-bit$_{\text{B}%
\rightarrow\text{A}}$.

\item Simultaneous, bidirectional transmission of two quantum bits:
q-bit$_{\text{A}\rightarrow\text{B}}$ and q-bit$_{\text{B}\rightarrow\text{A}%
}$.
\end{enumerate}

\item Analytical characterization of two--qubit gates according to their
capability to perform any of the above tasks (section \ref{sec4}).
\end{itemize}

\section{Definitions and basic facts}

This section is a prelude providing the definitions and notations that will be
used throughout the whole paper and reviews some facts concerning two-qubit
gates which will build the basis for our further results. We shall also define
the notion of majorization and collect some lemmas linked to it.

\subsection{Two-qubit gates\label{gates}}

Consider a system consisting of two two-dimensional subsystems (qubits), $A$
and $B$. The corresponding Hilbert spaces are $\mathcal{H}_{A\,}%
\approx\mathbb{C}^{2}$ and $\mathcal{H}_{B}\approx\mathbb{C}^{2}.$ The
compound Hilbert space is $\mathcal{H}_{AB}=\mathcal{H}_{A}\otimes
\mathcal{H}_{B}\approx\mathbb{C}^{2}\otimes\mathbb{C}^{2}$.

By a two-qubit gate $\mathcal{U}$ we understand a unitary operator acting on
$\mathcal{H}_{AB}$. By choosing the global phase appropriately we can always
consider such a unitary to be an element of the group $su(4,\mathbb{C})$. We
speak of a \emph{local} two-qubit gate whenever we can write $\mathcal{U}%
=U_{A}\otimes V_{B}$ where $U_{A}$ and $V_{B}$ are unitary operators acting
only on $\mathcal{H}_{A}$, $\mathcal{H}_{B}$ respectively. Again we can
restrict ourselves to local unitaries being elements of $su(2,\mathbb{C}%
)\otimes su(2,\mathbb{C})$. Non-local gates are then trivially two-qubit gates
which cannot be written as $U_{A}\otimes V_{B}$.

With just the help of these two definitions we can already divide the set of
non-local gates into equivalence classes. Two two-qubit gates $\mathcal{U}$
and $\widetilde{\mathcal{U}}$ \ are said to be \emph{locally equivalent} if
there exist local unitaries $U_{A}\otimes V_{B}$ and $\widetilde{U}_{A}%
\otimes\widetilde{V}_{B}$ such that $\mathcal{U=}U_{A}\otimes V_{B}%
\widetilde{\mathcal{U}} $ $\widetilde{U}_{A}\otimes\widetilde{V}_{B}$. A
useful decomposition of a general two-qubit gate developed in \cite{kha} and
\cite{kra} admits to further characterize these equivalence classes enabling
us to easily decide whether two gates are locally equivalent:

\begin{lemma}
\cite{kha},\cite{kra} For any two-qubit gate $\mathcal{U}$ \ there exist local
unitaries $U_{A}\otimes V_{B}$ and $\widetilde{U}_{A}\otimes\widetilde{V}_{B}$
and a self-adjoint operator of the form $H=\underset{k=1}{\overset{3}{\sum}%
}\alpha_{k}\sigma_{k}\otimes\sigma_{k}$ such that $\mathcal{U=}\widetilde
{U}_{A}\otimes\widetilde{V}_{B}e^{-iH}U_{A}\otimes V_{B}$. \label{canform}
\end{lemma}

Here the $\sigma_{k}$s denote the usual Pauli spin matrices. Note that the
real numbers $\alpha_{k}$ are not unique as long as we do not pose further
conditions on them. This is so for two reasons: Firstly operators of the type
$\pm\sigma_{k}\otimes\sigma_{k}$ are local and commute with $H$ so that we can
always extract such a local operator from the local parts in this
decomposition and include it in $H$. This alters the corresponding coefficient
$\alpha_{k}$ by $\pm\pi/2$. Secondly there are certain local transformations
of $H$ which conserve its form but permute the coefficients $\alpha_{k}$ and
change the sign of two of them. The local unitaries which cause such a
transformation are of the types $\pm i\sigma_{k}\otimes\mathbf{1}$ and $\pm
i\mathbf{1}\otimes\sigma_{k}$. Using this it can easily be checked that it is
always possible to bring $H$ to a form where its coefficients obey the
inequalities (see also \cite{kra})
\begin{equation}
\pi/4\geq\alpha_{1}\geq\alpha_{2}\geq\left|  \alpha_{3}\right|  \text{.}
\label{cond1}%
\end{equation}
Note that these conditions are an arbitrary choice and that it might be
necessary to relax them when we are looking for optimal simulation protocols.
We will come back to this point later on.

We call the decomposition of a two-qubit gate as given in lemma \ref{canform}
where the coefficients $\alpha_{k}$ fulfill (\ref{cond1}) its \emph{canonical
form}. The purely non-local unitary $e^{-iH}$ in this decomposition is termed
the \emph{interaction content} of the gate.

That the non-local characteristics of a two-qubit gate are determined by only
three real parameters is a remarkable result in view of the fact that a
general element of $su(4,\mathbb{C})$ is fixed by 15 independent parameters.
It might be mentioned here that while \cite{kha} provides a profound
Lie-algebraic basis for the decomposition in lemma \ref{canform}, \cite{kra}
gives a constructive proof which allows to determine the coefficients
$\alpha_{k}$ as well as the local unitaries for any given gate. Based on this
method we show in appendix \ref{appA} how to derive the $\alpha_{k}$ for a
given $\mathcal{U}$ without constructing the local unitaries.

A necessary and sufficient criterion for two gates to be locally equivalent is
now obviously that they have the same interaction content. By definition it is
also clear that any two-qubit gate is locally equivalent to its own
interaction content, a fact on which our results concerning simulation of
gates heavily rely.

For later use we mention here that self adjoint operators of the form
considered in lemma \ref{canform} are diagonal in the so called magic basis
\cite{hil} defined as%
\begin{equation}%
\begin{tabular}
[c]{ll}%
$\left|  \mathbf{1}\right\rangle =-\frac{i}{\sqrt{2}}\left(  \left|
01\right\rangle +\left|  10\right\rangle \right)  ,$ & $\left|  \mathbf{2}%
\right\rangle =\frac{1}{\sqrt{2}}\left(  \left|  00\right\rangle +\left|
11\right\rangle \right)  ,$\\
$\left|  \mathbf{3}\right\rangle =-\frac{i}{\sqrt{2}}\left(  \left|
00\right\rangle -\left|  11\right\rangle \right)  ,$ & $\left|  \mathbf{4}%
\right\rangle =\frac{1}{\sqrt{2}}\left(  \left|  01\right\rangle -\left|
10\right\rangle \right)  .$%
\end{tabular}
\ \label{mb}%
\end{equation}
Such that we have%
\begin{equation}
H=\underset{k=1}{\overset{3}{\sum}}\alpha_{k}\sigma_{k}\otimes\sigma
_{k}=\overset{4}{\underset{j=1}{\sum}}\lambda_{j}\left|  \mathbf{j}%
\right\rangle \left\langle \mathbf{j}\right|  \label{ham}%
\end{equation}
where the eigenvalues $\lambda_{j}$ follow from the $\alpha_{k}$ by%
\begin{equation}%
\begin{tabular}
[c]{ll}%
$\lambda_{1}=\alpha_{1}+\alpha_{2}-\alpha_{3},$ & $\lambda_{2}=\alpha
_{1}-\alpha_{2}+\alpha_{3},$\\
$\lambda_{3}=-\alpha_{1}+\alpha_{2}+\alpha_{3},$ & $\lambda_{4}=-\alpha
_{1}-\alpha_{2}-\alpha_{3}$.
\end{tabular}
\ \label{transf}%
\end{equation}
In terms of the $\lambda_{j}$ conditions (\ref{cond1}) read $3\pi/4\geq
\lambda_{1}\geq\lambda_{2}\geq\lambda_{3}\geq\lambda_{4}\geq-3\pi/4$. Note
also that the $\lambda_{j}$s sum up to zero (i.e. $H$ is traceless) such that
the corresponding unitary $\mathcal{U}=\exp(-iH_{\vec{\lambda}})$ is an
element of the special unitary group as we have required. In the following we
will characterize the interaction-content of non-local gates either by the
three-vector $\vec{\alpha}=(\alpha_{1},\alpha_{2},\alpha_{3})$ or by the four
vector $\vec{\lambda}=(\lambda_{1},\lambda_{2},\lambda_{3},\lambda_{4})$
freely switching between the representations. For operators like in
(\ref{ham}) we write $H_{\vec{\alpha}}$ or $H_{\vec{\lambda}}$ and for the
corresponding unitary $\mathcal{U}_{\vec{\alpha}}$ or $\mathcal{U}%
_{\vec{\lambda}}$.

\subsection{Majorization\label{maj}}

The relation of majorization emerged as a powerful tool in the issue of
simulation as well as in other fields of quantum information theory. From an
intuitive perspective it simply makes a precise statement out of a vague
notion that the components of a vector $\vec{x}$ are ''less spread out'' or
''more equal'' than are the components of a vector $\vec{y}$.

\begin{definition}
Let $\vec{x}=\left(  x_{1},...,x_{n}\right)  $ and $\vec{y}=(y_{1},...,y_{n})$
be real vectors whose components are ordered nonincresingly. Then we say that
''$\vec{x}$ majorizes $\vec{y}$'' and write $\vec{x}\succ\vec{y}$ if%
\begin{align*}
\overset{k}{\underset{i=1}{\sum}}x_{i}  &  \geq\overset{k}{\underset{i=1}%
{\sum}}y_{i}\qquad k=1,...,n-1\\
\overset{n}{\underset{i=1}{\sum}}x_{i}  &  =\overset{n}{\underset{i=1}{\sum}%
}y_{i}%
\end{align*}
\end{definition}

A central result in the theory of majorization is the following:

\begin{lemma}
\cite{mar} Let $x$ and $y$ be defined as before. Then $\vec{x}\succ\vec{y}$
iff there exists a doubly stochastic\footnote{A matrix is called doubly
stochatic if its entries are all nonnegative and each row and column adds up
to one.} $n\times n$ matrix $Q$ such that $\vec{y}=Q\vec{x}$.\label{hardy}
\end{lemma}

We will use two facts related to doubly stochastic matrices:

\begin{itemize}
\item The first one is called Birkhoff's theorem and states that the set of
doubly stochastic matrices is the convex hull of the permutation matrices.
Therefore we can write $Q=\sum p_{i}P_{i}$ (the $p_{i}\geq0$ summing up to one
and $P_{i}$ being permutation matrices) for any doubly stochastic matrix $Q$.

\item If we take the so called Hadamard product of a real orthogonal matrix
$O$ with itself i.e. square it componentwise (written symbolically as $O\circ
O$) then we get a special type of doubly stochastic matrix called
orthostochastic matrix.
\end{itemize}

Later on we will use this relation to compare 4-vectors $(\vec{\lambda}%
,\vec{\mu},\vec{\nu}...)$ of the kind introduced in the foregoing section. In
related works (\cite{ben},\cite{vic}) it has already turned out to be
convenient to have at hand an equivalent relation for the corresponding
3-vectors $(\vec{\alpha},\vec{\beta},\vec{\gamma}...)$ called the
s(pecial)-majorization relation. Let $\vec{\alpha}$ and $\vec{\beta}$ be two
real and nonincreasingly ordered 3-vectors. Then $\vec{\alpha}$ s-majorizes
$\vec{\beta}$ ($\vec{\alpha}\succ_{s}\vec{\beta}$) if%
\begin{align}
\alpha_{1}  &  \geq\beta_{1}\nonumber\\
\alpha_{1}+\alpha_{2}-\alpha_{3}  &  \geq\beta_{1}+\beta_{2}-\beta
_{3}\label{smajdef}\\
\alpha_{1}+\alpha_{2}+\alpha_{3}  &  \geq\beta_{1}+\beta_{2}+\beta
_{3}\nonumber
\end{align}
Now let $\vec{\lambda}$ and $\vec{\mu}$ be the 4-vectors related to
$\vec{\alpha}$ and $\vec{\beta}$ respectively via (\ref{transf}). Then it is
easily verified that $\vec{\lambda}\succ$ $\vec{\mu}$ iff $\vec{\alpha}$
$\succ_{s}$ $\vec{\beta}$.

The s-majorization relation can be extended to non-ordered vectors as follows.
Given a vector $\vec{\alpha}=(\alpha_{1},\alpha_{2},\alpha_{3})$, we construct
a new ``s-ordered'' vector $\vec{\alpha}^{s}=(\alpha_{1}^{s},\alpha_{2}%
^{s},\alpha_{3}^{s})$, $\alpha_{1}^{s}\geq\alpha_{2}^{s}\geq\left|  \alpha
_{3}^{s}\right|  $ by first nonincreasingly reordering the modulus of the
components $\alpha_{i}$, and by then giving $\alpha_{3}^{s}$ the sign of the
product $\alpha_{1}\alpha_{2}\alpha_{3}$. Then for any pair of vectors
$\vec{\alpha}$ and $\vec{\beta}$, $\vec{\alpha}\succ_{s}\vec{\beta}$ denotes
the set of inequalities (\ref{smajdef}) applied to $\vec{\alpha}^{s}$ and
$\vec{\beta}^{s}$. We note also that according to the above discussion a gate
$\mathcal{U}_{\vec{\alpha}}$ ($\vec{\alpha}$ being an \emph{arbitrary}
3-vector) is locally equivalent to the gate $\mathcal{U}_{\vec{\alpha}^{s}}$
corresponding to the s-ordered form of $\vec{\alpha}$.

\section{Interaction costs of gate simulation and partial order of gates}

\label{sec3}

The main result (theorem 1) in \cite{vid} permits to assess the interaction
cost (as defined in \cite{vid}) for simulating a two-qubit gate using any
given interaction Hamiltonian and fast local unitaries analytically after
performing a simple optimization. The proof in \cite{vid} is based on results
developed in the areas of quantum control \cite{kha} and quantum information
(\cite{kra},\cite{ben},\cite{due}). Here we give an alternative proof relying
only on the tools introduced so far. We do this by giving a necessary and
sufficient condition for the existence of a simulation protocol. Before we
state and prove this result we will introduce the problem of simulating a gate
(see \cite{ben},\cite{woc} for a more general discussions) and describe some
simplifications that can be assumed in this context.

\subsection{Setting of gate simulation and basic assumptions}

Simulating a desired two-qubit gate $\mathcal{U}$ using a given interaction
described by a Hamiltonian $H$ \footnote{Throughout the paper we put $\hbar=1$
and consider time to be dimensionless.} and arbitrary local unitary
transformations means to specify a series of local unitaries $\left\{
U_{1}\otimes V_{1},\ldots,U_{n}\otimes V_{n}\right\}  $ and of time intervals
$\left\{  t_{1},\ldots,t_{n}\right\}  $ such that%
\begin{align}
\mathcal{U}  &  \mathcal{=}\left(  U_{n}\otimes V_{n}\right)  e^{-iHt_{n}%
}\left(  U_{n-1}\otimes V_{n-1}\right)  e^{-iHt_{n-1}}\cdots
\nonumber\label{prot}\\
&  \cdots e_{1}^{-iHt_{2}}\left(  U_{1}\otimes V_{1}\right)  e^{-iHt_{1}%
}\left(  U_{0}\otimes V_{0}\right)  \text{.}%
\end{align}
Such a partition of a gate $\mathcal{U}$ equals a list of instructions like:
``Perform transformation $U_{0}$ and $V_{0}$ on qubit A and B respectively.
Then let them interact according to $H$ for a time $t_{1}$. Perform $U_{1}$
and $V_{1}$. Let them interact for $t_{2}$.$\cdot\cdot\cdot$Finally perform
$U_{n}$ and $V_{n}$.'' Following this protocol one would then effectively
perform the gate $\mathcal{U}$ on the two qubits no matter what their initial
state was.

Posing the problem of finding such a simulation protocol naturally evokes
other questions: Is there always a solution? How much time will it take to
perform a possible simulation protocol? What is the minimal time of
simulation? Do we have to allow for infinitesimal time steps? In case we can
restrict on taking finite time steps, how many of them will suffice? In the
following we will give an answer to all of them.

To do so we adopt two simplifications. At first we employ a physical
idealization namely the \emph{fast control limit} which is well justified in
most of the proposed settings for quantum information processing. It states
that the control operations - in our case the local unitary transformations -
can be executed in times where the natural evolution - here the interaction of
the qubits - has no considerable effect on the system's state. In other words
local manipulations and interactions have to take place on significantly
different time scales. That is what we assume and what allows us to define the
simulation time simply as $t_{S}=\underset{i=1}{\overset{n}{\sum}}t_{i}$
implying that the local transformations in (\ref{prot}) take effectively
\emph{no} time. We term the minimal time $t_{S}$ such that we can find a
simulation protocol its ``interaction cost'' ($\mathcal{C}_{H}(\mathcal{U)}$)
because it actually measures the \emph{time of interaction} required to
perform the gate.

The second simplification is of pure mathematical nature and concerns the
system's Hamiltonian $H$. Based on results of \cite{ben},\cite{due} we use
that although a general two-qubit Hamiltonian has the form $H=c_{0}%
\mathbf{1}\otimes\mathbf{1+}\overset{3}{\underset{i=1}{\sum}}a_{i}\sigma
_{i}\otimes\nolinebreak \mathbf{1}+\overset{3}{\underset{j=1}{\sum}}%
b_{j}\mathbf{1}\otimes\sigma_{j}+\underset{i,j=1}{\overset{3}{\sum}}%
c_{ij}\sigma_{i}\otimes\sigma_{j}$ we can restrict ourselves to much simpler
Hamiltonians $H_{\vec{\lambda}}$ (or equivalently $H_{\vec{\alpha}}$) as given
in (\ref{ham}). This is due to the fact that for any general Hamiltonian there
exists a Hamiltonian $H_{\vec{\lambda}}$, called its \emph{canonical form},
and efficient protocols for simulating the evolution according to the latter
in terms of the first. By an efficient simulation protocol we mean that we can
obtain the evolution $e^{-iH_{\vec{\lambda}}t}$ for any time $t$ by using $H$
for the same period of time $t$. (Note that such a simulation involves
infinitesimal time steps, see \cite{ben}.) For the purpose of simulation these
Hamiltonians are equivalent in the sense that both are equally effective in
simulating other Hamiltonians or gates.

\subsection{Necessary and sufficient condition for gate simulation}

We are now ready to give a necessary and sufficient condition for the
existence of a simulation protocol.

\begin{Result}
\label{Res1}Given a two-qubit gate $\mathcal{U}$ having an interaction content
$\mathcal{U}_{\vec{\beta}}$ and a Hamiltonian $H$ having a canonical form
$H_{\vec{\alpha}}$ there exists a simulation protocol of type (\ref{prot})
consuming a total time $t_{S}\geq0$ iff a vector $\vec{n}=(n_{1},n_{2},n_{3})$
of integers exists such that $\vec{\beta}_{\vec{n}}=\vec{\beta}+\pi/2\vec{n}$
satisfies%
\begin{equation}
\vec{\beta}_{\vec{n}}\prec_{s}\vec{\alpha}t_{S}\text{.} \label{result1}%
\end{equation}
\end{Result}

\begin{proof}
We first show that this is a necessary condition. According to the above
discussion a simulation protocol for $\mathcal{U}$ using $H$ for a time $t$ is
equivalent to a protocol for $\mathcal{U}_{\vec{\beta}}$ using $H_{\vec
{\alpha}}$ for the same time $t$. Moreover we can assume that the protocol we
have consists entirely of infinitesimal time steps $\delta t$ since any finite
time step can be decomposed into infinitesimal ones. Then (\ref{prot}) reads
as:%
\[%
\begin{tabular}
[c]{l}%
$\mathcal{U}_{_{\vec{\beta}}}\mathcal{=}\left(  U_{n}\otimes V_{n}\right)
e^{-iH_{\vec{\alpha}}\delta t}\left(  U_{n-1}\otimes V_{n-1}\right)  \cdots$\\
\multicolumn{1}{r}{$\cdots e^{-iH_{\vec{\alpha}}\delta t}\left(  U_{i}\otimes
V_{i}\right)  e^{-iH_{\vec{\alpha}}\delta t}\cdots\left(  U_{1}\otimes
V_{1}\right)  e^{-iH_{\vec{\alpha}}\delta t}\left(  U_{0}\otimes V_{0}\right)
$}%
\end{tabular}
\]
Let us assume that at a time $0\leq t\leq t_{S}$ we perform the $i^{th}$
intermediate local transformation having then attained an effective
transformation $\mathcal{U}_{t}=\left(  U_{i}\otimes V_{i}\right)
e^{-iH_{\vec{\alpha}}\delta t}\cdots\left(  U_{1}\otimes V_{1}\right)
e^{-iH_{\vec{\alpha}}\delta t}\left(  U_{0}\otimes V_{0}\right)  $. Since
$\mathcal{U}_{t}$ is itself a gate, we can decompose it as $\mathcal{U}%
_{t}=U_{t}\otimes V_{t}\mathcal{U}_{\vec{\gamma}_{t}}\widetilde{U}_{t}%
\otimes\widetilde{V}_{t}$ where $\mathcal{U}_{\vec{\gamma}_{t}}=e^{-iH_{\vec
{\gamma}_{t}}}$ is the interaction content of $\mathcal{U}_{t}$. The index $t$
indicates the time dependence of all these unitaries.

To determine how $\vec{\gamma}_{t}$ varies with $t$ we take the next
infinitesimal time step $e^{-iH_{\vec{\alpha}}\delta t}$ in the protocol and
get%
\begin{align*}
e^{-iH_{\vec{\alpha}}\delta t}\mathcal{U}_{t} &  =e^{-iH_{\vec{\alpha}}\delta
t}U_{t}\otimes V_{t}\mathcal{U}_{\vec{\gamma}_{t}}\widetilde{U}_{t}%
\otimes\widetilde{V}_{t}\\
&  =U_{t+\delta t}\otimes V_{t+\delta t}\mathcal{U}_{\vec{\gamma}_{t+\delta
t}}\widetilde{U}_{t+\delta t}\otimes\widetilde{V}_{t+\delta t}\text{.}%
\end{align*}
For convenience we change here to the 4-vector representation [as defined in
(\ref{transf})]. Denote by $\vec{\lambda},\vec{\nu},\vec{\xi}$ the vectors
corresponding to $\vec{\alpha},\vec{\gamma}_{t},\vec{\gamma}_{t+\delta t}$
respectively. After local transformations the last identity can be written as%
\begin{equation}
e^{-iH_{\vec{\lambda}}\delta t}U\otimes V\mathcal{U}_{\vec{\nu}}=W\otimes
X\mathcal{U}_{\vec{\xi}}Y\otimes Z\label{infinitesimal}%
\end{equation}
where $W\otimes X$ and $Y\otimes Z$ are appropriately defined local unitaries
and all time indices are omitted. The right hand side of (\ref{infinitesimal})
is a decomposition of the left hand side, but we do not require this to be the
canonical form as defined in section \ref{gates}. We therefore have the
possibility to put further conditions on the unitaries in this decomposition.

If we multiply from the left by $U^{\dagger}\otimes V^{\dagger}$ and sandwich
this equation between $\left|  \mathbf{k}\right\rangle $, one of the magic
states, we find%
\begin{equation}
\left\langle \psi_{k}\right|  e^{-iH_{\vec{\lambda}}\delta t}\left|  \psi
_{k}\right\rangle e^{-i\nu_{k}}=\left\langle \psi_{k}\right|  W\otimes
X\mathcal{U}_{\vec{\xi}}Y\otimes Z\left|  \mathbf{k}\right\rangle
\label{diagelem}%
\end{equation}
where $\left|  \psi_{k}\right\rangle :=U\otimes V\left|  \mathbf{k}%
\right\rangle $. In order to have equality for $\delta t=0$ we make use of the
above mentioned freedom and require for this case $W\otimes X=U\otimes
V,Y\otimes Z=\mathbf{1\otimes1}$ and $\vec{\xi}=\vec{\nu}$.

For infinitesimal $\delta t$ we can thus expand%
\begin{align*}
\left\langle \psi_{k}\right|  W\otimes X  & =\left\langle \mathbf{k}\right|
+\left\langle \delta\mathbf{k}^{\bot}\right|  \\
Y\otimes Z\left|  \mathbf{k}\right\rangle  & =\left|  \mathbf{k}\right\rangle
+\left|  \delta\mathbf{\bar{k}}^{\bot}\right\rangle \\
\vec{\xi}  & =\vec{\nu}+\delta\vec{\nu}.
\end{align*}
where we may assume $\left\langle \delta\mathbf{k}^{\bot}\right|
\mathbf{k}\rangle=\langle k\left|  \delta\mathbf{\bar{k}}^{\bot}\right\rangle
=0$. Combining everything in (\ref{diagelem}) and collecting terms up to first
order we find%
\[
\left\langle \psi_{k}\right|  H_{\vec{\lambda}}\left|  \psi_{k}\right\rangle
\delta t=\delta\nu_{k}%
\]
which has to hold for all $k$.

Let us now take a closer look at the diagonal elements $\left\langle \psi
_{k}\right|  H_{\vec{\lambda}}\left|  \psi_{k}\right\rangle $. With regard to
the definition $\left|  \psi_{k}\right\rangle $ and now again including the
time dependence of $U_{t}\otimes V_{t}$ we have $\left\langle \psi_{k}\right|
H_{\vec{\lambda}}\left|  \psi_{k}\right\rangle =\left\langle \mathbf{k}%
\right|  \left(  U_{t}\otimes V_{t}\right)  ^{\dagger}H_{\vec{\lambda}}\left(
U_{t}\otimes V_{t}\right)  \left|  \mathbf{k}\right\rangle $. In the magic
basis local unitaries take on the form of real orthogonal matrices [$\left(
U_{t}\otimes V_{t}\right)  ^{\dagger}\rightarrow O(t)$] and the Hamiltonian
gets diagonal [$H_{\vec{\lambda}}\rightarrow D_{\vec{\lambda}}:=$%
diag$(\vec{\lambda})$]. Therefore $\delta\nu_{k}=\delta t(OD_{\vec{\lambda}%
}O^{T})_{kk}=\delta t[(O\circ O)\vec{\lambda}]_{k}$ where $(O\circ O)$ denotes
the Hadamard product of the real orthogonal matrix $O(t)$ with itself.
Defining $Q(t):=O(t)\circ O(t)$ we can write compactly%
\begin{equation}
\frac{\delta\vec{\nu}}{\delta t}=Q(t)\vec{\lambda}\text{.}\label{xi}%
\end{equation}

Recall that $\delta\vec{\nu}$ is the variation of the interaction content at
some intermediate time $0\leq t\leq t_{S}$ in our simulation protocol. The
overall interaction content $\vec{\nu}(t_{S})$ is found by integrating
(\ref{xi}) from $0$ to $t_{S}$. As initial condition we have $\vec{\nu
}(0)=\vec{0}$ since our simulation protocol starts from the identity having no
interaction content. We then find%
\[
\vec{\nu}(t_{S})=\underset{0}{\overset{t_{S}}{\int}}Q(t)dt\vec{\lambda}%
=S\vec{\lambda}t_{S}%
\]
where $S:=1/t_{S}\underset{0}{\overset{t_{S}}{\int}}Q(t)dt$ is again a doubly
stochastic matrix. To see this observe $\overset{4}{\underset{j=1}{\sum}%
}S_{jk}=1/t_{S}\underset{0}{\overset{t_{S}}{\int}}\overset{4}{\underset
{j=1}{\sum}}Q(t)_{jk}dt=1/t_{S}\underset{0}{\overset{t_{S}}{\int}}1dt=1$. The
same holds for summation over $k$.

With lemma \ref{hardy} we can state that $\vec{\nu}(t_{S})\prec\vec{\lambda
}t_{S}$ or switching again to the 3-vector representation $\vec{\gamma}%
(t_{S})\prec_{s}\vec{\alpha}t_{S}$\ [see the definitions preceding equation
(\ref{infinitesimal})]. Remember that our basic assumption was that we have a
simulation protocol for a gate $\mathcal{U}=U\otimes V\mathcal{U}_{\vec{\beta
}}\widetilde{U}\otimes\widetilde{V}$. However, by means of $\vec{\gamma}%
(t_{S})$ we can find a - possibly different - decomposition since
$\mathcal{U}=\mathcal{U}_{t_{S}}=U_{t_{S}}\otimes V_{t_{S}}\mathcal{U}%
_{\vec{\gamma}_{t_{S}}}\widetilde{U}_{t_{S}}\otimes\widetilde{V}_{t_{S}}$.
From the discussion in section \ref{gates} we know that the vectors
$\vec{\beta}$ and $\vec{\gamma}_{t_{S}}$ have to be related via the local
operations specified there. There are two operations that can be done to alter
$\overrightarrow{\beta}$: (i) add multiples of $\pi/2$ to its components, i.e.
build $\vec{\beta}_{\vec{n}}=\vec{\beta}+\pi/2\vec{n}$ for a vector $\vec
{n}=(n_{1},n_{2},n_{3})$, and (ii) permute and simultaneously change the sign
of two components, which can be expressed easily by multiplication with an
appropriate matrix $P$. Therefore we must have $\vec{\gamma}(t_{S}%
)=P\vec{\beta}_{\vec{n}}\prec_{s}\vec{\alpha}t_{S}$ for some $P$ and $\vec{n}%
$. Recalling the definition of s-ordering of vectors [see (\ref{smajdef}) and
the remarks there] we find $\left(  P\vec{\beta}_{\vec{n}}\right)
_{s}=\left(  \vec{\beta}_{\vec{n}}\right)  _{s}$ and therefore $\vec{\beta
}_{\vec{n}}\prec_{s}\vec{\alpha}t_{S}$.

We now turn to the second part of our proof and show sufficiency. Since this
has already been proven in \cite{vic} we will just sketch this proof. Let
$\vec{\mu}$ and $\vec{\lambda}$ be the 4-vectors corresponding to $\vec{\beta
}_{\vec{n}}$ and $\vec{\alpha}$. Then (\ref{result1}) reads as $\vec{\mu}%
\prec\vec{\lambda}t_{S}$ and it follows by Brikhoff's theorem (see section
\ref{maj}) that we can write $\vec{\mu}=\underset{i=1}{\overset{n}{\sum}}%
p_{i}P_{i}\vec{\lambda}t_{S}=\underset{i=1}{\overset{n}{\sum}}P_{i}%
\vec{\lambda}t_{i}$ where we defined $t_{i}=p_{i}t_{S}$. Using that each of
the $4!=24$ permutations $P_{i}$ of the magic states $\left\{  \left|
\mathbf{j}\right\rangle \right\}  $ can be performed through appropriate local
unitaries $U_{i}\otimes V_{i}$ we have%
\begin{align*}
\mathcal{U}_{\vec{\beta}_{\vec{n}}} &  =e^{-iH_{\vec{\mu}}}=\exp\left(
-i\underset{i=1}{\overset{n}{\sum}}H_{P_{i}\vec{\lambda}t_{i}}\right)  \\
&  =\exp\left(  -i\underset{i=1}{\overset{n}{\sum}}U_{i}\otimes V_{i}%
H_{\vec{\lambda}}U_{i}^{\dagger}\otimes V_{i}^{\dagger}t_{i}\right)  \\
&  =\overset{n}{\underset{i=1}{\prod}}U_{i}\otimes V_{i}e^{-iH_{\vec{\lambda}%
}t_{i}}U_{i}^{\dagger}\otimes V_{i}^{\dagger}\text{.}%
\end{align*}
For the last line we took into account that $\left[  U_{i}\otimes V_{i}%
H_{\vec{\lambda}}U_{i}^{\dagger}\otimes V_{i}^{\dagger},U_{j}\otimes
V_{j}H_{\vec{\lambda}}U_{j}^{\dagger}\otimes V_{j}^{\dagger}\right]  =0$
$\forall i,j$ since the local transformations involved only permute the
eigenvectors of $H_{\vec{\lambda}}$. The last line provides clearly a proper
simulation protocol for $\mathcal{U}_{\vec{\beta}_{\vec{n}}}$ and - by
applying appropriate local unitaries at the beginning and at the end - for all
locally equivalent gates (including $\mathcal{U}_{\vec{\beta}}$).
\end{proof}

We remark here that \cite{vic} shows how to find explicitly the probability
distribution $\left\{  p_{i}\right\}  $ and permutations $\left\{
P_{i}\right\}  $ which determine the time steps $\left\{  t_{i}\right\}  $ and
the local unitaries $\left\{  U_{i}\otimes V_{i}\right\}  $. There also the
maximal number n of evolution steps sufficient in any simulation protocol was
determined. It turned out to be three for time optimal protocols.

This condition for the simulation of gates is an analogue to the one
established in \cite{ben} for efficient Hamiltonian simulation. Such a
correspondence was, in principle, only expected for infinitesimal gates. It is
remarkable that it extends in such a tight analogy to finite gates. The main
difference is that here we have to include all different decompositions of the
gate under consideration by allowing for variations $\vec{\beta}_{\vec{n}%
}=\vec{\beta}+\pi/2\vec{n}$. There is no analog to this in the case of
Hamiltonian simulation. The reason for this is that here we have to
accommodate the periodicity properties of unitary operators while in the
setting of Hamiltonian simulation we deal with a linear space of Hermitian operators.

\subsection{Interaction costs\label{intcosts}}

To finally assess the interaction cost $\mathcal{C}_{H}(\mathcal{U)}$ - i.e.
the minimal time to simulate $\mathcal{U}$ using $H$ and local unitaries as
defined in \cite{vid}- we just have to optimize condition (\ref{result1}) with
respect to both $t_{S}$ and $\vec{n}$. Doing so we reproduce the main result
of \cite{vid}:

\begin{Result}
\label{Res2}The interaction cost $\mathcal{C}_{H}(\mathcal{U)}$ is the minimal
value of $t_{S}\geq0$ such that either $\vec{\beta}_{(0,0,0)}\prec_{s}%
\vec{\alpha}t_{S}$ or $\vec{\beta}_{(-1,0,0)}\prec_{s}\vec{\alpha}t_{S}$ holds.
\end{Result}

\begin{proof}
This is equivalent to result \ref{Res1} under the restriction that it suffices
to look at $\vec{n}$ being $(0,0,0)$ or $(-1,0,0)$ to find the smallest
$t_{S}$. This is because in case $\vec{n}$ is not one of these two vectors we
can show that either $\vec{\beta}_{(0,0,0)}\prec_{s}\vec{\beta}_{\vec{n}}$ or
$\vec{\beta}_{(-1,0,0)}\prec_{s}\vec{\beta}_{\vec{n}}$. For the minimal time
$t_{S}$ such that $\vec{\beta}_{\vec{n}}\prec_{s}\vec{\alpha}t_{S}$ for a
given $\vec{n}$ we therefore essentially have either $\vec{\beta}%
_{(0,0,0)}\prec_{s}\vec{\alpha}t_{S}$ or $\vec{\beta}_{(-1,0,0)}\prec_{s}%
\vec{\alpha}t_{S}$ for the same time $t_{S}$. Obviously letting $\vec{n}$ be
$(0,0,0)$ or $(-1,0,0)$ will make for \emph{at least} the same minimal time.
The optimization for $(0,0,0)$ or $(-1,0,0)$ cannot be avoided since in
general $\vec{\beta}_{(0,0,0)}$ and $\vec{\beta}_{(-1,0,0)}$ are incomparable
according to the s-majorization relation. To show that either $\vec{\beta
}_{(0,0,0)}\prec_{s}\vec{\beta}_{\vec{n}}$ or $\vec{\beta}_{(-1,0,0)}\prec
_{s}\vec{\beta}_{\vec{n}}$ for all $\vec{n}$ different than $(0,0,0)$ or
$(-1,0,0)$ we distinguish two cases. (i) First we look at vectors $\vec{n}$
having at least one component $\left|  n_{j}\right|  >1$. Since the components
of $\vec{\beta}$ have to fulfill (\ref{cond1}) the maximal component of the
reordered form of $\vec{\beta}_{\vec{n}}$ (see section \ref{gates}) is at
least $3\pi/4$. We then have $\vec{\beta}_{\vec{n}}\succ_{s}(3\pi/4,0,0)$ and
this last vector clearly s-majorizes both $\vec{\beta}_{(0,0,0)}$ and
$\vec{\beta}_{(-1,0,0)}$. (ii) The vectors $\vec{n}$ satisfying $\left|
n_{j}\right|  \leq1\forall j$ have to be checked case by case. We find
$\vec{\beta}_{(-1,0,0)}\prec_{s}\vec{\beta}_{\vec{n}}$ for $\vec{n}\in\left\{
(-1,-1,-1),(0,1,0),(0,0,-1),(0,0,1)\right\}  $ and $\vec{\beta}_{(0,0,0)}%
\prec_{s}\vec{\beta}_{\vec{n}}$ for the remaining $\vec{n}$.
\end{proof}

Let us formulate result \ref{Res2} as a kind of recipe. In order to time
optimally perform a gate $\mathcal{U}$ using an interaction described by a
Hamiltonian $H$ together with arbitrary local unitaries proceed as follows:

\begin{enumerate}
\item Determine $\vec{\beta}$ characterizing the interaction content of
$\mathcal{U}$ following \cite{kra} (see also appendix \ref{appA}). Using
\cite{ben},\cite{due} compute the canonical form of $H$ to get $\vec{\alpha}$.

\item Test whether $\vec{\beta}$ or $\vec{\beta}_{(-1,0,0)}$ is s-majorized by
$\vec{\alpha}t_{S}$ for a smaller time $t_{S}$.

\item For the vector yielding the better result as well as for $\vec{\alpha}$
compute the corresponding 4-vectors $\vec{\mu}$ and $\vec{\lambda}$
respectively. Following \cite{vic} find the permutations $P_{i}$ and
probabilities $p_{i}$ $(i=1,2,3)$ such that $\vec{\mu}=\underset{i=1}%
{\overset{3}{\sum}}p_{i}P_{i}\vec{\lambda}t_{S}$.

\item The $p_{i}$ determine the time steps $t_{i}$ and the $P_{i}$ give the
local unitaries to be applied in between. This provides a simulation protocol
for $\mathcal{U}_{\vec{\mu}}$ using $H_{\vec{\lambda}}$ for at most 3 finite
time steps.

\item Simulate the evolutions according to $H_{\vec{\lambda}}$ by using the
Hamiltonian $H$ for the same period of time following \cite{ben}. Apply
appropriate local unitaries (determined using \cite{kra}) in the beginning and
at the end of the overall simulation to effectively perform $\mathcal{U}$.
\end{enumerate}

We now discuss certain special cases for which some of the above points can be
dropped or get simpler.

\begin{itemize}
\item In case the Hamiltonian we use describes solely pure interaction, that
is to say is of the form $H=\underset{i,j=1}{\overset{3}{\sum}}c_{ij}%
\sigma_{i}\otimes\sigma_{j}$ without any local parts, we can attain its
canonical form by a local transformation $H_{\vec{\alpha}}=U\otimes
VHU^{\dagger}\otimes V^{\dagger}$ (see \cite{ben}). Since $e^{-iH_{\vec
{\alpha}}t_{i}}=U\otimes Ve^{-iHt_{i}}U^{\dagger}\otimes V^{\dagger}$ we do
not have to employ infinitesimal simulations (as required in step 5.) and the
simulation protocol will only contain 3 finite time steps.

\item In case the interaction content of the desired gate is characterized by
a vector $\vec{\beta}=(\beta_{1},\beta_{2},\beta_{3})$ satisfying $\vec{\beta
}\prec_{s}\vec{\beta}_{(-1,0,0)}$ we can skip the optimization (step 2.) and
state directly: The interaction cost $\mathcal{C}_{H}(\mathcal{U)}$ is the
minimal value of $t_{S}$ such that $\vec{\beta}\prec_{s}\vec{\alpha}t_{S}$.
The condition on $\vec{\beta}$ for $\vec{\beta}\prec_{s}\vec{\beta}%
_{(-1,0,0)}$ to be true is $\beta_{1}+\left|  \beta_{3}\right|  \leq\pi/4$. To
see this we have to apply the inequalities (\ref{smajdef}) defining the
s-majorization to $\vec{\beta}$ and $\vec{\beta}_{(-1,0,0)}^{s}=(\pi
/2-\beta_{1},\beta_{2},-\beta_{3})$, the s-ordered version of $\vec{\beta
}_{(-1,0,0)}$ (see section \ref{gates}). We find:%
\begin{align*}
\beta_{1}  &  \leq\pi/2-\beta_{1}\\
\beta_{1}+\beta_{2}\mp\beta_{3}  &  \leq\pi/2-\beta_{1}+\beta_{2}\pm\beta_{3}%
\end{align*}
The first inequality is fulfilled trivially since $\beta_{1}\leq\pi/4$ in any
case. The last two inequalities are equivalent to $\beta_{1}+\left|  \beta
_{3}\right|  \leq\pi/4$ and this is what we claimed. The reverse $\vec{\beta
}_{(-1,0,0)}\prec_{s}\vec{\beta}$ is never true because the first inequality
is violated for any $\beta_{1}$. In all the other cases where $\vec{\beta}$
and $\vec{\beta}_{(-1,0,0)}$ are incomparable it will depend on the
Hamiltonian which of the two vectors yields the optimal time.
\end{itemize}

\subsection{Interaction costs of basic gates\label{basicgates}}

As an illustration we shall give here explicitly the interaction costs for
three specific gates (CNOT, D(ouble)CNOT, SWAP) and for the whole class of
controlled-$U$ gates. We choose these ones not only because they play a
prominent role in quantum information but also due to their role as
``landmarks'' in the set of two-qubit gates as we will show in the next
section. Let us list them here by first giving their definition in terms of
their action on the computational basis $\left\{  \left|  i,j\right\rangle
\right\}  _{i,j=0}^{1}$, then characterizing their interaction content by the
corresponding vector $\vec{\beta}$ and finally assessing the interaction costs
pursuant to a general Hamiltonian $H$ with canonical form $H_{\vec{\alpha}}$.

\subsubsection{CNOT gate and controlled-$U$ gates}

The C(ontrolled)NOT gate is the prototypical two-qubit quantum logic gate. Its
action is defined compactly as $\left|  i\rangle_{A}\otimes|j\right\rangle
_{B}\rightarrow\left|  i\rangle_{A}\otimes|i\oplus j\right\rangle _{B}$ where
$\oplus$ denotes addition modulo 2. That is, it flips the second (target)
qubit iff the first (control) qubit is in state $\left|  1\right\rangle $. Let
us denote the CNOT gate by $\mathcal{U}_{CNOT}^{AB}$ where the first
superscript indicates the control and the second the target qubit. In appendix
\ref{appA} we show that the interaction content of this gate is given by
$\vec{\beta}=\pi/4(1,0,0)$. Therefore the CNOT belongs to the special class of
gates where we can skip the optimization in result \ref{Res2} and go straight
ahead to majorization in order to determine the interaction cost. Requiring
$\vec{\beta}\prec_{s}\vec{\alpha}t_{S}$ is equivalent to:
\begin{align*}
\pi/4  &  \leq\alpha_{1}t_{S}\\
\pi/4  &  \leq(\alpha_{1}+\alpha_{2}\pm\alpha_{3})t_{S}%
\end{align*}
Clearly the first inequality yields the tighter bound. The interaction cost
for simulating a CNOT is $\mathcal{C}_{H}(CNOT)=\frac{\pi}{4}\frac{1}%
{\alpha_{1}}$.

The CNOT is a representative of the general class of controlled-$U$ gates.
These gates apply a unitary operation on the target qubit iff the control
qubit is in state $\left|  1\right\rangle $. Thus they have the form%
\[
\mathcal{U}_{ctrl-U}\mathcal{=}\left|  0\right\rangle \left\langle 0\right|
\otimes\mathbf{1+}\left|  1\right\rangle \left\langle 1\right|  \otimes
U\text{\textbf{.}}
\]
In appendix \ref{appA} we show that the interaction content of a
controlled-$U$ gate is always described by $\vec{\beta}=(\beta,0,0)$ where
$\beta$ is fixed by the eigenvalues of $U$. The interaction cost to simulate
such a gate is $\mathcal{C}_{H}(\mathcal{U}_{ctrl-U})=\frac{\beta}{\alpha_{1}%
}$.

\subsubsection{DCNOT gate}

The D(ouble)CNOT gate is the concatenation of two CNOTs in the following way
$\mathcal{U}_{DCNOT}^{AB}\mathcal{=U}_{CNOT}^{BA}\mathcal{U}_{CNOT}^{AB}$ and
its action on the computational basis can be described as $\left|
i\rangle_{A}\otimes|j\right\rangle _{B}\rightarrow\left|  j\rangle_{A}%
\otimes|i\oplus j\right\rangle _{B}$. This gate was introduced in \cite{coll}
as an intermediate gate between the CNOT and the SWAP. In the following we
will emphasise the special role of the DCNOT gate. Its interaction content is
described by $\vec{\beta}=\pi/4(1,1,0)$ such that the DCNOT falls as well
under the class of gates where we do not have to care about the optimization.
For the interaction cost we find $\mathcal{C}_{H}(DCNOT)=\frac{\pi}{4}%
\frac{2}{\alpha_{1}+\alpha_{2}-\left|  \alpha_{3}\right|  }$.

\subsubsection{SWAP gate}

The SWAP gate is the unique gate having the effect to exchange the states of
two qubits i.e. transforming $\left|  i\rangle_{A}\otimes|j\right\rangle
_{B}\rightarrow\left|  j\rangle_{A}\otimes|i\right\rangle _{B}$. It is well
known that $\mathcal{U}_{SWAP}\mathcal{=U}_{CNOT}^{AB}\mathcal{U}_{CNOT}%
^{BA}\mathcal{U}_{CNOT}^{AB}$ and regarding the two other gates not very
surprising that its interaction content is $\vec{\beta}=\pi/4(1,1,1)$. Once
more recalling conditions (\ref{cond1}) we can say that this is maximal. Now
the optimization can not be avoided. We find $\vec{\beta}_{(-1,0,0)}^{s}%
=\pi/4(1,1,-1)$ and it turns to be optimal to simulate $\vec{\beta}$ $\left(
\vec{\beta}_{(-1,0,0)}^{s}\right)  $ if $\alpha_{3}>0$ $\left(  \alpha
_{3}<0\right)  $. In case $\alpha_{3}=0$ the interaction costs are equal for
both alternatives. In any case we find the interaction costs $\mathcal{C}%
_{H}(SWAP)=\frac{\pi}{4}\frac{3}{\alpha_{1}+\alpha_{2}+\left|  \alpha
_{3}\right|  }$.

\subsection{Order of gates}

What we see by these examples and what was to be expected is that the
interaction costs depend strongly on the interaction resource - i.e. the
Hamiltonian - we have at our disposal. But once the interaction is fixed the
notion of interaction cost induces an order in the set of gates allowing us to
compare the ``non-locality'' of two gates in terms of the resources needed to
perform them. Of course this order is always relative to the Hamiltonian and
may change when we choose another one. For example if we use the Ising
interaction $\sigma_{1}\otimes\sigma_{1}$ we find the CNOT to be less
non-local than the DCNOT and this one in turn to be less non-local than the
SWAP. On the contrary with the exchange interaction $\sigma_{1}\otimes
\sigma_{1}+\sigma_{2}\otimes\sigma_{2}+\sigma_{3}\otimes\sigma_{3}$ at hand
the SWAP is less time consuming than the DCNOT and in this sense less
non-local. However in a restricted region of the set of two-qubit gates this
order is absolute in that it does not depend on the interaction Hamiltonian.
We will first define this order properly and then state and prove this result:

We say gate $\mathcal{U}$ is \emph{more non-local} than gate $\mathcal{V}$,
and write $\mathcal{V}\leq\mathcal{U}$, when for all interactions $H$ the
interaction cost of $\mathcal{U}$ is never smaller than that of $\mathcal{V}$,%

\[
\mathcal{V}\leq\mathcal{U\quad}\equiv\quad\mathcal{C}_{H}(\mathcal{V}%
)\leq\mathcal{C}_{H}(\mathcal{U})\quad\forall H\text{.}
\]

\begin{Result}
\label{Res3}Let $\mathcal{U}$ and $\mathcal{V}$ be two two-qubit gates with
corresponding ordered vectors $\vec{\beta}_{\mathcal{U}}$ and $\vec{\beta
}_{\mathcal{V}}$ such that in both cases the restriction $\beta_{1}+\left|
\beta_{3}\right|  \leq\pi/4$ holds. Then gate $\mathcal{U}$ is more non-local
than gate $\mathcal{V}$ if and only if $\vec{\beta}_{\mathcal{V}}\prec_{s}%
\vec{\beta}_{\mathcal{U}}$.
\end{Result}

\begin{proof}
Since both vectors $\vec{\beta}$ satisfy $\beta_{1}+\left|  \beta_{3}\right|
\leq\pi/4$ the interaction costs $\mathcal{C}_{H}(\mathcal{V})$ and
$\mathcal{C}_{H}(\mathcal{U})$ are given, respectively, by the smallest
$t_{\mathcal{U}},t_{\mathcal{V}}\geq0$ such that%
\begin{align*}
\vec{\beta}_{\mathcal{V}}  &  \prec_{s}\vec{\alpha}t_{\mathcal{V}},\\
\vec{\beta}_{\mathcal{U}}  &  \prec_{s}\vec{\alpha}t_{\mathcal{U}}.
\end{align*}
Suppose first $\mathcal{V}\leq\mathcal{U}$, that is, for any Hamiltonian $H$
we have $\mathcal{C}_{H}(\mathcal{V})\leq\mathcal{C}_{H}(\mathcal{U})$ and in
particular $\vec{\beta}_{\mathcal{V}}\prec_{s}\vec{\alpha}\mathcal{C}%
_{H}(\mathcal{U})$. If we rewrite this relation for the particular Hamiltonian
where $\vec{\alpha}=\vec{\beta}_{\mathcal{U}}$ and use that in this case
$\mathcal{C}_{H}(\mathcal{U})=1$ we find $\vec{\beta}_{\mathcal{V}}\prec
_{s}\vec{\beta}_{\mathcal{U}}$. This proves the direct implication. The
inverse follows right away by using the partial order property of
majorization. $\vec{\alpha}\mathcal{C}_{H}(\mathcal{U})\succ_{s}\vec{\beta
}_{\mathcal{U}}\succ_{s}\vec{\beta}_{\mathcal{V}}$ directly implies
$\mathcal{C}_{H}(\mathcal{U})\geq$ $\mathcal{C}_{H}(\mathcal{V})$ (see the
proof of result \ref{Res2}).
\end{proof}

Once more coming back to the problem of Hamiltonian simulation we mention that
the corresponding partial order there has been solved completely \cite{ben}.
The reason why the partial order established in result \ref{Res3} only holds
in the region of gates where $\beta_{1}+\left|  \beta_{3}\right|  \leq\pi/4$
is again that we have to deal here with the rather involved periodic structure
of $su(4)$. It is exactly this restricted region where we can evade this
difficulty by suppressing the otherwise essential optimization between
$\vec{\beta}_{(0,0,0)}$ and $\vec{\beta}_{(-1,0,0)}$ (step 2 in the recipe
given in section \ref{intcosts}).

\section{Transmission of information and classes of gates}

\label{sec4}

By now we analysed two-qubit gates in terms of the time expense they cause in
the context of simulation. There the main objective is to perform a given gate
on two qubits using a minimum time of interaction seen as a valuable resource.
The notion of interaction cost thereby obtained gave a measure for how
non-local (either relative to a specific interaction or absolute as in result
\ref{Res3}) a gate is. In this section we change the perspective. We now want
to prescribe the tasks a gate has to accomplish and ask how non-local it
therefore has to be. In this setting we consider the gate and its inherent
non-locality to be the valuable resource. The task we have in mind here is the
transmission of information in form of classical as well as quantum bits.

This section is organized as follows: First we motivate why the capability of
gates to transmit bits is a proper measure for their non-locality. After
having given some basic definitions, we collect a number of known results for
certain gates. Then we treat the problems of transmitting a cbit or a qubit in
one direction as well as all possible combinations of them in both directions
by using a two-qubit gate and determine the interaction content necessary to
do so. The subsequent discussion of the results will allow us to distinguish
various classes of gates differing in their capability for
quantum-communicational tasks which will give a characterization of the
non-locality of a gate as well .

\subsection{Transmission capability and non-local content of gates}

Non-local gates result physically from an interaction taken place between the
qubits by some means. Interaction between two physical systems conditions on
the other hand the transmission of information between them since after having
interacted (at least one of) the subsystem's states will have changed
depending on the states of \emph{both} subsystems as they were before the
interaction. Hence there must have been some kind of information exchange in
the process of interaction. It is therefore natural to ask whether we can
utilise a non-local gate to send (classical or quantum) information. The
amount of information we can transmit using a gate will give us then a
characterization of its degree of non-locality. A similar point of view was
captured in \cite{coll},\cite{eis} where the amount of classical and quantum
information necessary to implement a gate was adapted as a measure for its
non-local content.

What do we mean by the transmission of classical or quantum information?
Consider two parties Alice and Bob holding a qubit A and B respectively.
Assume further that somehow they manage to perform a gate $\mathcal{U}$ on
their qubits. Then we say that $\mathcal{U}$ allows for the \emph{transmission
of a classical-bit} from Alice to Bob (denoted by cbit$_{\text{A}%
\rightarrow\text{B}}$) if after the application of $\mathcal{U}$ Bob can
distinguish with probability 1 whether Alice's qubit was in $\left|
0\right\rangle $ or $\left|  1\right\rangle $. We speak of the
\emph{transmission of a quantum-bit} from Alice to Bob (qubit$_{\text{A}%
\rightarrow\text{B}}$) if under the action of $\mathcal{U}$ Bob's qubit takes
on the state of Alice's qubit.

Let us make some remarks here. (i) The essential difference between these two
effects of a gate is, that in the case of cbit$_{\text{A}\rightarrow\text{B}}$
we do not require superpositions of $\left|  0\right\rangle $ and $\left|
1\right\rangle $ to be transmitted faithfully whereas in the case
qubit$_{\text{A}\rightarrow\text{B}}$ we do. The possibility to send a qubit
trivially includes the one to transmit a cbit resembling the fact that quantum
information incorporates classical information. (ii) Without further
specifying $\mathcal{U}$ we can state directly that in case qubit$_{\text{A}%
\rightarrow\text{B}}$ Alice loses her state after sending it due to the
no-cloning-theorem. (iii) If Alice's qubit is maximally entangled to some
ancilla qubit on her side then the transmission qubit$_{\text{A}%
\rightarrow\text{B}}$ swaps the entanglement thus establishing a maximally
entangled pair of qubits (e-bit) between Alice and Bob. That is why the
authors of \cite{coll},\cite{eis} identified the capabilities of a gate to
send a qubit and to create an e-bit. Here we want to distinguish between the
actual \emph{creation of entanglement} without ancilla systems as treated in
\cite{kra} and \emph{entanglement swapping} by the transmission of a qubit.
This differentiation is essential for example in the case of a CNOT gate which
can be used to create an e-bit $\left[  \mathcal{U}_{CNOT}^{AB}\frac{1}%
{\sqrt{2}}(\left|  0\right\rangle +\left|  1\right\rangle )\otimes\left|
0\right\rangle =\frac{1}{\sqrt{2}}(\left|  00\right\rangle +\left|
11\right\rangle )\right]  $ but not to transmit a qubit as we will show in the following.

For the gates introduced in section \ref{basicgates} it is well known and easy
to see how they can be used to transmit bits. Regarding the definitions given
there the following is effortless verified:

\begin{itemize}
\item $\mathcal{U}_{CNOT}^{AB}\left|  i0\right\rangle =\left|  ii\right\rangle
,i=0,1$ and therefore the CNOT is sufficient to send a cbit from Alice to Bob.
Since Alice's qubit does not change a all under the action of this gate it is
impossible for her to send a qubit to Bob (see remark (ii) above). This is not
true if Alice and Bob share entanglement as an additional resource. See the
remark below.

\item $\mathcal{U}_{DCNOT}^{AB}\left|  \varphi0\right\rangle =\left|
0\varphi\right\rangle $ where $\left|  \varphi\right\rangle $ is an arbitrary
qubit state transmitted by the action of $\mathcal{U}_{DCNOT}^{AB}$. Moreover
we find $\mathcal{U}_{DCNOT}^{AB}\left|  \varphi1\right\rangle =\left|
1\rangle\otimes\sigma_{x}|\varphi\right\rangle $ telling us that Bob may send
at the same time a cbit to Alice under the condition that in case he sent
$\left|  1\right\rangle $ he flips his qubit after the transmission in order
to recover the correct state $\left|  \varphi\right\rangle $. Since he knows
what he sent, as we can assume, this requires no additional communication.

\item $\mathcal{U}_{SWAP}\left|  \varphi\psi\right\rangle =\left|  \psi
\varphi\right\rangle $ where $\left|  \varphi\right\rangle $ and $\left|
\psi\right\rangle $ are arbitrary states both being transmitted faithfully.
\end{itemize}

We can summarize this by the implications:%
\begin{align*}
\text{CNOT}  &  \rightarrow\text{cbit}_{\text{A}\rightarrow\text{B}}\\
\text{DCNOT}  &  \rightarrow\text{qubit}_{\text{A}\rightarrow\text{B}%
}\text{+cbit}_{\text{B}\rightarrow\text{A}}\\
\text{SWAP}  &  \rightarrow\text{qubit}_{\text{A}\rightarrow\text{B}%
}\text{+qubit}_{\text{B}\rightarrow\text{A}}%
\end{align*}
Obviously, due to the symmetry of the non-local content of two-qubit gates
under exchange of parties, the same expressions hold if we make the
substitutions A$\leftrightarrow$B. These relations hold strictly for the case
where the communicating parties have no ancilla systems and no prior
entanglement at hand, but have to be read as lower bounds on the capabilities
of these gates to transfer information if we allow for additional resources of
this kind. It is a central result in quantum information that the capacities
to transmit information can be increased if the parties possess shared
entanglement (e-bits) \cite{coll},\cite{eis}.

\subsection{Transmission of information in the context of gate simulation}

Assume now Alice and Bob want to send some given amount of information
(possibly in both directions) by using some fixed interaction described by a
Hamiltonian $H$ and arbitrary local transformations of their qubits. They
could do so by choosing appropriately one of the above gates providing the
necessary transmission capability and then simulate it according to the
results we derived so far. The interaction costs thereby incurred are given in
section \ref{basicgates}. But is this optimal? There might be gates which are
suitable for the same task but have an interaction content different from the
ones of CNOT, DCNOT or SWAP yielding smaller interaction costs. In the
following we want to single out which gate is both sufficient for a certain
transmission task and optimal in terms of interaction costs. We do this by
deriving necessary and sufficient conditions on the interaction content of a
gate to be capable for the transmission of a given amount of information. All
we have to do then is to find the gate which fulfills the appropriate
condition and causes the minimal interaction cost.

\subsubsection{cbit$_{\text{A}\rightarrow\text{B}}$}

Assume Alice encodes a classical bit into her qubit by preparing it in
$\left|  0\right\rangle $ or $\left|  1\right\rangle $ and Bob holds some
arbitrary state $\left|  \varphi\right\rangle $. Then the bit is by definition
transmitted if after an application of a gate $\mathcal{U}$ Bob's qubit takes
on a state $\left|  \psi\right\rangle $ or $\left|  \psi^{\perp}\right\rangle
$ (some state orthogonal to $\left|  \psi\right\rangle $) depending on whether
Alice sent ``0'' or ``1''. At the same time Alice's qubit may change
arbitrarily. The action of $\mathcal{U}$ we have to require is described by%
\begin{equation}%
\begin{array}
[c]{ccc}%
\left|  0\varphi\right\rangle  & \rightarrow & \left|  \chi\psi\right\rangle
\\
\left|  1\varphi\right\rangle  & \rightarrow & \left|  \widetilde{\chi}%
\psi^{\perp}\right\rangle \text{.}%
\end{array}
\label{cbittransm}%
\end{equation}
More precisely we can state: A necessary condition for a gate $\mathcal{U}$ to
be capable of transmitting a cbit is, that there exist states $\left|
\varphi\right\rangle ,\left|  \chi\right\rangle ,\left|  \widetilde{\chi
}\right\rangle ,\left|  \psi\right\rangle $ and $\left|  \psi^{\perp
}\right\rangle $ such that relations (\ref{cbittransm}) hold. Assume now that
this is indeed the case. What can we say about the interaction content of
$\mathcal{U}$? Since independent local transformations before and after the
application of $\mathcal{U}$ do not affect its interaction content, we can
look for unitaries fulfilling $Z\left|  \varphi\right\rangle =\left|
0\right\rangle ,Y\left|  \chi\right\rangle =\left|  0\right\rangle ,X\left|
\psi\right\rangle =\left|  0\right\rangle $ and $X\left|  \psi^{\perp
}\right\rangle =\left|  1\right\rangle $ and define $\mathcal{U}^{\prime
}=(X_{A}\otimes Y_{B})\mathcal{U}(\mathbf{1}_{A}\otimes Z_{B})$ having a
simpler action given by
\begin{equation}%
\begin{array}
[c]{ccc}%
\left|  00\right\rangle  & \rightarrow & \left|  00\right\rangle \\
\left|  10\right\rangle  & \rightarrow & \left|  \alpha1\right\rangle \text{.}%
\end{array}
\label{cbittransm2}%
\end{equation}
where $\left|  \alpha\right\rangle =Y\left|  \widetilde{\chi}\right\rangle $.
$\mathcal{U}^{\prime}$ and $\mathcal{U}$ are locally equivalent and therefore
have the same interaction content. To derive conditions on this interaction
content we apply $\mathcal{U}^{\prime}$ to the state $\varrho:=\frac{1}%
{2}\mathbf{1}_{A}\otimes\left|  0\right\rangle _{B}\left\langle 0\right|  $ -
transforming under the terms of (\ref{cbittransm2}) - and take the partial
trace with respect to system $A$:%
\begin{equation}
tr_{A}\left\{  \mathcal{U}^{\prime}\varrho\mathcal{U}^{\prime\dagger}\right\}
=\frac{1}{2}\text{tr}\left\{  \left|  00\right\rangle \left\langle 00\right|
+\left|  \alpha1\right\rangle \left\langle \alpha1\right|  \right\}
=\frac{1}{2}\mathbf{1}_{B}. \label{cbit1}%
\end{equation}
When we on the other hand assume a decomposition $\mathcal{U}^{\prime
}=(\widetilde{U}\otimes\widetilde{V})\mathcal{U}_{\vec{\beta}}(U\otimes V)$ we find%

\begin{equation}%
\begin{tabular}
[c]{l}%
$tr_{A}\left\{  \mathcal{U}^{\prime}\varrho\mathcal{U}^{\prime\dagger
}\right\}  =$\\
\multicolumn{1}{r}{$=\frac{1}{2}tr_{A}\left\{  \left[  \widetilde{V}%
_{B}\mathcal{U}_{\vec{\beta}}^{AB}V_{B}\right]  \mathbf{1}_{A}\mathbf{\otimes
}\left|  0\right\rangle _{B}\left\langle 0\right|  \left[  \widetilde{V}%
_{B}\mathcal{U}_{\vec{\beta}}^{AB}V_{B}\right]  ^{\dagger}\right\}  .$}%
\end{tabular}
\label{cbit2}%
\end{equation}
Equating the right hand sides of (\ref{cbit1}) and (\ref{cbit2}) and
multiplying from the left by $\widetilde{V}_{B}^{\dagger}$ and from the right
by $\widetilde{V}_{B}$ yields%
\[
\mathbf{1}_{B}=\text{tr}_{A}\left\{  \mathcal{U}_{\vec{\beta}}^{AB}%
\mathbf{1}_{A}\mathbf{\otimes}\left|  \omega\right\rangle _{B}\left\langle
\omega\right|  \mathcal{U}_{\vec{\beta}}^{AB\dagger}\right\}
\]
where we have abbreviated $V\left|  0\right\rangle =\left|  \omega
\right\rangle $. Expressing without loss of generality $\left|  \omega
\right\rangle =\cos(\omega)\left|  0\right\rangle +e^{-i\theta}\sin
(\omega)\left|  1\right\rangle $ one can work out the trace explicitly and
finds%
\begin{align}
\mathbf{1}_{B}  &  =\left(
\begin{array}
[c]{cc}%
1-a & b\\
b^{\ast} & 1+a
\end{array}
\right) \label{cbit3}\\
a  &  =\cos(2\omega)\cos(2\beta_{1})\cos(2\beta_{2})\nonumber\\
b  &  =\sin(2\omega)\cos(2\beta_{3})[\cos(\theta)\cos(2\beta_{2})+i\sin
(\theta)\sin(2\beta_{1})]\text{.}\nonumber
\end{align}
Let us stop here and consider what equation (\ref{cbit3}) tells us. The left
hand side was an immediate consequence of the necessary conditions on
$\mathcal{U}$ to properly transmit a cbit while the right hand side results
from the general ansatz $\mathcal{U}^{\prime}=(\widetilde{U}\otimes
\widetilde{V})\mathcal{U}_{\vec{\beta}}(U\otimes V)$ where the unitary $V$
contains the parameters $\omega,\theta$ and $\vec{\beta}=(\beta_{1},\beta
_{2},\beta_{3})$ characterizes the interaction content $\mathcal{U}%
_{\vec{\beta}}$. Equation (\ref{cbit3}) thus puts certain conditions on the
parameters in the decomposition of $\mathcal{U}^{\prime}$. Obviously we have
to require $a=b=0$. This in turn is fulfilled in various cases, for example
whenever two of the coefficients $\beta_{k}=\pi/4$, the third being arbitrary.
However it is also easy to see that there are solutions, where only \emph{one}
of the coefficients $\beta_{k}=\pi/4$.\ In this case we have to choose either
$\omega$ or $\theta$ appropriately. This puts conditions on the state $\left|
\varphi\right\rangle $ in (\ref{cbittransm}) denoting the input state Bob has
to choose in order to properly receive the cbit Alice aims to send him. Three
solutions of this kind are for example given by $\left\{  \beta_{1}%
=\pi/4,\omega=0\right\}  ,\left\{  \beta_{2}=\pi/4,\omega=0\right\}  $ and
$\left\{  \beta_{3}=\pi/4,\omega=\pi/4\right\}  $ where in each case the
remaining parameters can be chosen arbitrarily. All in all we have shown that
it is a \emph{necessary} condition for the transmission of a cbit to have at
least one of the coefficients $\beta_{k}$ equal to $\pi/4$ and without loss of
generality we can always require this to be $\beta_{1}$.

To be systematic we should now continue and show, that any gate characterized
by a vector $\vec{\beta}=(\pi/4,\beta_{2},\beta_{3})$ is also
\emph{sufficient} for this task. But at this point we will not do so for two
reasons. Firstly we already know that an interaction content $\vec{\beta}%
=(\pi/4,0,0)$ is sufficient to transmit a cbit because this basically fixes a
CNOT or any gate locally equivalent to a CNOT. Secondly we find $(\pi
/4,0,0)\prec_{s}(\pi/4,\beta_{2},\beta_{3})$ for all $0\leq\left|  \beta
_{3}\right|  \leq\beta_{2}\leq\pi/4$ and therefore $\mathcal{C}_{H}%
(CNOT)\leq\mathcal{C}_{H}(\mathcal{U}_{(\pi/4,\beta_{2},\beta_{3})})$ for all
$H$. Thus looking for gates other than ones out of the CNOT-class has no
advantage in terms of interaction costs. Let us state this as

\begin{Result}
The cheapest (time optimal) way to transmit a cbit using some given
interaction is to simulate a CNOT gate. The interaction cost is $\mathcal{C}%
_{H}($cbit$_{\text{A}\rightarrow\text{B}})=\frac{\pi}{4}\frac{1}{\alpha_{1}}$.
\end{Result}

The following results will show that the transmission capability scales up
with the coefficients $\beta_{k}$ becoming bigger. Just by continuity it
follows then right away that any gate having an interaction content
$\vec{\beta}=(\pi/4,\beta_{2},\beta_{3})$ is also sufficient to tranmit at
least a cbit.

\subsubsection{cbit$_{\text{A}\rightarrow\text{B}}$ and cbit$_{\text{B}%
\rightarrow\text{A}}$}

Again let Alice encode a logical bit into her qubit as $\left|  0\right\rangle
$ or $\left|  1\right\rangle $. Further assume Bob wants to send ``0'' and
therefore prepares $\left|  0\right\rangle $. To properly transmit their two
messages they have to find a gate, which transforms the states like%
\begin{equation}%
\begin{array}
[c]{ccc}%
\left|  00\right\rangle  & \rightarrow & \left|  \varphi\chi\right\rangle \\
\left|  10\right\rangle  & \rightarrow & \left|  \psi\chi^{\perp}\right\rangle
\text{.}%
\end{array}
\label{cbitsdef1}%
\end{equation}
To detect the messages being sent to him, Bob has to measure the observable
$\sigma_{\chi}=\left|  \chi\right\rangle \left\langle \chi\right|  -\left|
\chi^{\perp}\right\rangle \left\langle \chi^{\perp}\right|  $. Conversely
Alice has to measure $\sigma_{\varphi}$ or $\sigma_{\psi}$ (defined similarly)
depending on whether her message was ``0'' or ``1''. Consider now the same
situation but let Bob's message be ``1''. The same reasoning as before yields%
\begin{equation}%
\begin{array}
[c]{ccc}%
\left|  01\right\rangle  & \rightarrow & \left|  \varphi^{\perp}%
\omega\right\rangle \\
\left|  11\right\rangle  & \rightarrow & \left|  \psi^{\perp}\omega^{\perp
}\right\rangle \text{.}%
\end{array}
\label{cbitsdef2}%
\end{equation}
Now Bob has to measure $\sigma_{\omega}$. The transformation behaviour
characterized so far lacks of one essential condition: it is not unitary.
Unitary transformations map an orthonormal basis into another one and this is
so long not fulfilled, since f.e. $\left\langle \varphi\chi\right|
\psi^{\perp}\omega^{\perp}\rangle\neq0$. Imposing that the vectors on the
right hand side of (\ref{cbitsdef1}) and (\ref{cbitsdef2}) build again a basis
one finds four possible cases: (i) $\left\langle \varphi\right|  \psi^{\bot
}\rangle=0$ and $\left\langle \psi\right|  \varphi^{\bot}\rangle=0$, (ii)
$\left\langle \chi\right|  \omega^{\bot}\rangle=0$ and $\left\langle
\omega\right|  \chi^{\bot}\rangle=0$, (iii) $\left\langle \varphi\right|
\psi^{\bot}\rangle=0$ and $\left\langle \omega\right|  \chi^{\bot}\rangle=0$
and (iv) $\left\langle \psi\right|  \varphi^{\bot}\rangle=0$ and $\left\langle
\chi\right|  \omega^{\bot}\rangle=0$. The last two cases are more restrictive
than (i) and (ii) since there the states of \emph{both} qubits have to meet
certain conditions. We are however interested to stay as less restrictive as
possible so that we are going to focus on (i) in which case we have to require
$\left|  \psi\right\rangle =e^{-i\alpha}\left|  \varphi\right\rangle $ and
$\left|  \psi^{\bot}\right\rangle =e^{-i\beta}\left|  \varphi^{\bot
}\right\rangle $. Let us summarize what we have found so far:%
\begin{align*}
\left|  00\right\rangle  &  \rightarrow\left|  \varphi\chi\right\rangle \\
\left|  10\right\rangle  &  \rightarrow e^{-i\alpha}\left|  \varphi\chi
^{\perp}\right\rangle \\
\left|  01\right\rangle  &  \rightarrow\left|  \varphi^{\perp}\omega
\right\rangle \\
\left|  11\right\rangle  &  \rightarrow e^{-i\beta}\left|  \varphi^{\perp
}\omega^{\perp}\right\rangle \text{.}%
\end{align*}
Including the phases into $\left|  \chi^{\bot}\right\rangle $ and $\left|
\omega^{\bot}\right\rangle $ and again adjusting the axes by local
transformations to cleanse the notation (as we did for the cbit$_{\text{A}%
\rightarrow\text{B}}$-problem) we can write equivalently%
\begin{equation}%
\begin{array}
[c]{ccc}%
\left|  00\right\rangle  & \rightarrow & \left|  00\right\rangle \\
\left|  10\right\rangle  & \rightarrow & \left|  01\right\rangle \\
\left|  01\right\rangle  & \rightarrow & \left|  1\omega\right\rangle \\
\left|  11\right\rangle  & \rightarrow & \left|  1\omega^{\perp}\right\rangle
\text{.}%
\end{array}
\label{cbits1}%
\end{equation}
We can see that Bob has to measure a different observable depending on what he
sent. For the case (ii) above we would find similar transformations but then
being Alice the one who has to adapt her observable. Therefore case (i) gets
identical with (ii), if we let Alice and Bob exchange their names which in
turn cannot have any relevance for the interaction content of the gate they
use. Or more mathematically: (i) can be transformed into (ii) by conjugating
the gate with the SWAP and this does not alter the interaction content.

We can now parametrize $\left|  \omega\right\rangle =\cos(\omega)\left|
0\right\rangle +e^{-i\theta}\sin(\omega)\left|  1\right\rangle $ and $\left|
\omega^{\bot}\right\rangle =e^{-i\eta}(-\sin(\omega)\left|  0\right\rangle
+e^{-i\theta}\cos(\omega)\left|  1\right\rangle )$ and determine the
interaction content of the gate%
\begin{align}
\mathcal{U}(\eta,\theta,\omega)  &  =e^{-i\pi/4}e^{i(\eta+\theta)/4}%
\times\nonumber\\
&  \times\left(
\begin{array}
[c]{cccc}%
e^{-i(\eta+\theta)}\cos(\omega) & 0 & e^{-i\theta}\sin(\omega) & 0\\
-e^{-i\eta}\sin(\omega) & 0 & \cos(\omega) & 0\\
0 & 1 & 0 & 0\\
0 & 0 & 0 & 1
\end{array}
\right)  \label{cbits}%
\end{align}
written in the computational basis $\left\{  \left|  11\right\rangle ,\left|
10\right\rangle ,\left|  01\right\rangle ,\left|  00\right\rangle \right\}  $
in this order. The global phase assures $\mathcal{U}(\eta,\theta,\omega)$
being a \emph{special }unitary operator. Following appendix \ref{appA} one
finds for the vector $\vec{\beta}=(\beta_{1},\beta_{2},\beta_{3})$
characterizing the interaction content $\mathcal{U}_{\vec{\beta}}$ of
$\mathcal{U}(\eta,\theta,\omega)$%
\begin{align*}
\beta_{1}  &  =\pi/4\\
\beta_{2}  &  =\pi/4\\
\beta_{3}  &  =\pi/4-\vartheta
\end{align*}
where $\vartheta$ is a solution to $\tan^{2}(2\vartheta)=\sec^{2}\left(
\frac{\eta+\theta}{2}\right)  \sec^{2}(\omega)-1$. $\vartheta$ therefore
parametrizes a family of gates, of which each element has the desired
capability to transmit cbit$_{\text{A}\leftrightarrow\text{B}}$.

Note especially that the DCNOT $\left[  \vec{\beta}=(\pi/4,\pi/4,0)\right]  $
and the SWAP $\left[  \vec{\beta}=(\pi/4,\pi/4,\pi/4)\right]  $ belong to this
family as we should expect according to the discussion in section
\ref{basicgates}. These gates are attained for the choice $\vartheta=\pi/4$
and $\vartheta=0$ respectively. In terms of $(\eta,\theta,\omega)$ this
corresponds f.e. to set $\left(  \eta=\pi,\theta=0,\omega=\pi/2\right)  $ and
$\left(  \eta+\theta=0,\omega=0\right)  $ for the DCNOT and the SWAP
respectively yielding the expected result when inserted in (\ref{cbits}).

If we want to tranmit the cbits using some given interaction we can freely
choose the parameter $\vartheta\ $out of $\left[  0,\pi/2\right]  $ in order
to keep down the interaction costs. Let us present the optimal choice in

\begin{Result}
\label{Res5}The cheapest (time optimal) way to transmit cbits in both
directions using some given interaction is to simulate a gate holding an
interaction content $\vec{\beta}=\frac{\pi}{4}(1,1,\frac{2\alpha_{3}}%
{\alpha_{1}+\alpha_{2}})$. The corresponding interaction cost is
$\mathcal{C}_{H}($cbit$_{\text{A}\leftrightarrow\text{B}})=\frac{\pi}%
{4}\frac{2}{\alpha_{1}+\alpha_{2}}$.
\end{Result}

\begin{proof}
Define $b:=1/2-2/\pi\cdot\vartheta$ and parametrize $\vec{\beta}%
(\vartheta)=\vec{\beta}(b)=\frac{\pi}{4}(1,1,2b)$. We have to find
$b\in\left[  -1/2,1/2\right]  $ and $t_{S}\geq0$ such that either $\vec{\beta
}(b)\prec_{s}\vec{\alpha}t_{s}$ or $\vec{\beta}_{(-1,0,0)}(b)\prec_{s}%
\vec{\alpha}t_{s}$ holds and $t_{S}$ is minimal. First note that $\vec{\beta
}_{(-1,0,0)}^{s}(b)=\vec{\beta}(-b)$. The optimization with respect to $b$
therefore includes the one with respect to $\vec{\beta}$ and $\vec{\beta
}_{(-1,0,0)}$. The minimal time such that $\vec{\beta}(b)\prec_{s}\vec{\alpha
}t_{s}$ is fulfilled is given by $t_{\min}(b)=\max\left\{  \frac{\pi}%
{4}\frac{1}{\alpha_{1}},\frac{\pi}{2}\frac{1-b}{\alpha_{1}+\alpha_{2}%
-\alpha_{3}},\frac{\pi}{2}\frac{1+b}{\alpha_{1}+\alpha_{2}+\alpha_{3}%
}\right\}  $. Optimization with respect to $b$ yields the interaction cost
$\mathcal{C}_{H}($cbit$_{\text{A}\leftrightarrow\text{B}})=\underset
{b\in\left[  -1/2,1/2\right]  }{\min}\left[  t_{\min}(b)\right]
=\underset{b\in\left[  -1/2,1/2\right]  }{\min}\left[  \max\left\{  \frac{\pi
}{4}\frac{1}{\alpha_{1}},\frac{\pi}{2}\frac{1-b}{\alpha_{1}+\alpha_{2}%
-\alpha_{3}},\frac{\pi}{2}\frac{1+b}{\alpha_{1}+\alpha_{2}+\alpha_{3}%
}\right\}  \right]  $. This is an exercise in linear optimization which has to
be solved under the condition $\pi/4\geq\alpha_{1}\geq\alpha_{2}\geq\left|
\alpha_{3}\right|  $. An elementary calculation yields $\mathcal{C}_{H}%
($cbit$_{\text{A}\leftrightarrow\text{B}})=\frac{\pi}{2}\frac{1}{\alpha
_{1}+\alpha_{2}}$ for $b=\frac{\alpha_{3}}{\alpha_{1}+\alpha_{2}}$.
\end{proof}

\subsubsection{qubit$_{\text{A}\rightarrow\text{B}}$ and (qubit$_{\text{A}%
\rightarrow\text{B}}$ and cbit$_{\text{B}\rightarrow\text{A}}$)}

To reliably transmit a qubit we have to require%
\begin{align*}
\left|  00\right\rangle  &  \rightarrow\left|  \varphi\chi\right\rangle \\
\left|  10\right\rangle  &  \rightarrow\left|  \varphi\chi^{\perp
}\right\rangle \text{.}%
\end{align*}
The remaining vectors $\left|  01\right\rangle $ and $\left|  11\right\rangle
$ may transform arbitrarily but have to stay orthogonal to both among
themselves and with respect to $\left|  \varphi\chi\right\rangle $ and
$\left|  \varphi\chi^{\perp}\right\rangle $. The least restrictive choice
yields similar to the foregoing section%
\begin{align*}
\left|  01\right\rangle  &  \rightarrow\left|  \varphi^{\perp}\omega
\right\rangle \\
\left|  11\right\rangle  &  \rightarrow\left|  \varphi^{\perp}\omega^{\perp
}\right\rangle \text{.}%
\end{align*}
Without loss of generality we can identify $\left|  \varphi\right\rangle
=\left|  0\right\rangle ,\left|  \varphi^{\perp}\right\rangle =\left|
1\right\rangle ,\left|  \chi\right\rangle =\left|  0\right\rangle $ and
$\left|  \chi^{\perp}\right\rangle =\left|  1\right\rangle $ ending up with
the same gate (\ref{cbits1}) as for cbit$_{\text{A}\leftrightarrow\text{B}}$.
The optimal interaction content and cost to send a qubit$_{\text{A}%
\rightarrow\text{B}}$ is therefore the same as in result \ref{Res5}.

Regarding the transformations given in (\ref{cbits1}) it is obvious that this
gate is also capable to send at the same time a cbit$_{\text{B}\rightarrow
\text{A}}$. To do so Bob encodes his bit into $\left|  0\right\rangle $ or
$\left|  1\right\rangle $. Applying $\mathcal{U}$ sends the bit to Alice. The
qubit Bob gets from Alice comes in faithfully if Bob sent ``0''. In the other
case he has to recover the qubit by a local transformation obeying $V\left|
\omega\right\rangle =\left|  0\right\rangle $ and $V\left|  \omega^{\perp
}\right\rangle =\left|  1\right\rangle $. An interaction content $\vec{\beta
}=(\pi/4,\pi/4,\pi/4-\vartheta)$ is therefore sufficient for the transmission
qubit$_{\text{A}\rightarrow\text{B}}$ and cbit$_{\text{B}\rightarrow\text{A}}%
$. This is also necessary since any interaction content showing less than
$\pi/4$ in the first two entries is not sufficient to send a qubit$_{\text{A}%
\rightarrow\text{B}}$. Again we can refer to the values given in result 5 for
the optimal interaction content and cost.

\subsubsection{qubit$_{\text{A}\leftrightarrow\text{B}}$}

This problem is trivial since the exchange of the two quantum states
completely fixes the transformation of the basis states and therefore also the
gate. The SWAP is the only gate providing the required action. Interaction
content and cost are given in section \ref{basicgates}.

\subsection{Classes of gates}

Let us summarize the results of the foregoing sections in the following table:%

\begin{widetext}
\begin{center}%

\begin{tabular}
[c]{r|ccc|lll|c}%
\multicolumn{4}{r|}{interaction content} & \multicolumn{3}{c|}{transmission
capability} & interaction cost\\\hline\hline
& $\beta_{1}$ & $\beta_{2}$ & $\beta_{3}$ & $\text{cbit}_{\text{A}%
\rightarrow\text{B}}$ & qubit$_{\text{A}\rightarrow\text{B}}$\&$\text{cbit}%
_{\text{B}\rightarrow\text{A}}$ & $\text{qubit}_{\text{A}\leftrightarrow
\text{B}}$ & \multicolumn{1}{|l}{$\mathcal{C}_{H}(\mathcal{U})$}%
\\\cline{2-4}\cline{2-8}%
\multicolumn{1}{r|}{controlled$\text{-}U$} & $x$ & $0$ & $0$ &
\multicolumn{1}{|c}{$\times$} & \multicolumn{1}{c}{$\times$} &
\multicolumn{1}{c|}{$\times$} & \multicolumn{1}{|l}{$\frac{x}{\alpha_{1}}$}\\
\multicolumn{1}{r|}{$\text{CNOT}$} & $\pi/4$ & $0$ & $0$ &
\multicolumn{1}{|c}{$\checkmark$} & \multicolumn{1}{c}{$\times$} &
\multicolumn{1}{c|}{$\times$} & \multicolumn{1}{|l}{$\frac{\pi}{4}%
\frac{1}{\alpha_{1}}$}\\
I & $\pi/4$ & $y$ & $z$ & \multicolumn{1}{|c}{$\checkmark$} &
\multicolumn{1}{c}{$\times$} & \multicolumn{1}{c|}{$\times$} &
\multicolumn{1}{|l}{$\geq\frac{\pi}{4}\frac{1}{\alpha_{1}}$}\\
\multicolumn{1}{r|}{$\text{DCNOT}$} & $\pi/4$ & $\pi/4$ & $0$ &
\multicolumn{1}{|c}{$\checkmark$} & \multicolumn{1}{c}{$\checkmark$} &
\multicolumn{1}{c|}{$\times$} & \multicolumn{1}{|l}{$\frac{\pi}{4}%
\frac{2}{\alpha_{1}+\alpha_{2}-\left|  \alpha_{3}\right|  }$}\\
\multicolumn{1}{r|}{II} & $\pi/4$ & $\pi/4$ & $z$ &
\multicolumn{1}{|c}{$\checkmark$} & \multicolumn{1}{c}{$\checkmark$} &
\multicolumn{1}{c|}{$\times$} & \multicolumn{1}{|l}{$\frac{\pi}{4}%
\frac{2}{\alpha_{1}+\alpha_{2}}$ for $z=\frac{\pi}{4}\frac{2\alpha_{3}}%
{\alpha_{1}+\alpha_{2}}$}\\
\multicolumn{1}{r|}{$\text{SWAP}$} & $\pi/4$ & $\pi/4$ & $\pi/4$ &
\multicolumn{1}{|c}{$\checkmark$} & \multicolumn{1}{c}{$\checkmark$} &
\multicolumn{1}{c|}{$\checkmark$} & \multicolumn{1}{|l}{$\frac{\pi}{4}%
\frac{3}{\alpha_{1}+\alpha_{2}+\left|  \alpha_{3}\right|  }$}%
\end{tabular}%

\end{center}
\end{widetext}%

What we can see, is that the capability of a gate to transmit information
increases when the coefficients $\beta_{k}$ characterizing its interaction
content approach their maximal values $\pi/4$. Especially when one of them
takes on this maximum value, the corresponding gate acquires a new feature.
The special gates CNOT, DCNOT, and SWAP (and all their local equivalents) mark
these thresholds and that is why we announced them being ``landmarks'' in the
set of two-qubit gates. This allows us to distinguish four classes of gates
differing in their transmission capability: (i) gates with $\pi/4>\beta
_{1}\geq\beta_{2}\geq\left|  \beta_{3}\right|  $ (no transmission capability),
(ii) CNOT and type I, (iii) DCNOT and type II and (iv) SWAP. This
classification endows the coefficients $\beta_{k}$ with physical significance
and therefore complements earlier work, where a gate's interaction content
$\mathcal{U}_{\vec{\beta}}$ was associated with its capability to create
entanglement \cite{kra}.

\section{Conclusions}

In this work we addressed the problem of simulating two-qubit gates using some
given interaction and local unitary transformations in the fast control limit.
For this to be possible we presented a necessary and sufficient condition
linking the gate, the Hamiltonian\ characterizing the interaction and the
total time of simulation. Optimization with respect to time gave a measure
$\mathcal{C}_{H}(\mathcal{U})$ - termed interaction cost - for how costly such
a simulation in terms of time of interaction is and thereby recovered a result
already attained in \cite{vid}. The interaction cost has been computed for
various gates and was shown to induce a partial order in a region of the set
of two-qubit gates thus establishing a meaningful notion of and measure for
the non-locality of a gate.

To give an application as well as a supplementation of these results we then
turned to the problem of transmitting information between two parties using
two-qubit gates. Necessary and sufficient conditions on gates were established
to be capable of transferring classical and quantum bits in all combinations
and directions. This allowed us to compute explicitly the interaction costs
for these tasks. Beyond it the transmission capability of a gate provided a
classification of two-qubit gates.

All results derived here concern two-qubit systems. All the underlying
problems can naturally be extended to higher dimensional systems and therefore
it would be desirable to generalize the results. The main obstacle to do so is
that in higher dimensions there is no decomposition like in ( \ref{canform})
for a general unitary operator.

\section{Acknowledgments}

K. H. would like to thank Barbara Kraus for kind and generous help and the
referee of this paper for a hint to simplify the proof of result \ref{Res1}.
We thank C.H. Bennett, A. Harrow, D. W. Leung and J. A. Smolin for
communications about their results on the use of bipartite Hamiltonians to
communicate information \cite{ben2}. This work was supported by the European
Community project EQUIP (contract IST-1999-11053) and by the National Science
Foundation of USA, grant No. EIA-0086038.

\appendix     

\section{Interaction content of non-local gates}

\label{appA}

In lemma \ref{canform} we presented a decomposition for two-qubit gates of the
form $\mathcal{U=}\widetilde{U}_{A}\otimes\widetilde{V}_{B}e^{-iH}U_{A}\otimes
V_{B}$ where $H=\exp(\overset{3}{\underset{k=1}{\sum}}\alpha_{k}\sigma
_{k}\otimes\sigma_{k})$. Here we demonstrate a method based on \cite{kra} to
determine the $\alpha_{k}$ for a general given $\mathcal{U}$.

In section \ref{gates} we gave an alternative representation of $H$ in terms
of its eigenvalues $\lambda_{k}$. The method actually admits to compute the
$\lambda_{k}$s and relies on the following two observations: (i) Hamiltonians
of the special form considered here are diagonal in the magic basis as we have
already shown in section \ref{gates}. (ii) Local unitaries are real in the
magic basis \cite{hil}. Especially they become real orthogonal matrices since
of course they stay to be unitary. This fact resembles the homomorphism
$su(2,\mathbb{C})\otimes su(2,\mathbb{C})\simeq SO(4,\mathbb{R)}$ \cite{gil}
becoming manifest in the magic basis. Using these two facts the decomposition
takes on the form $\mathcal{U}=\widetilde{O}DO$ when written in the magic
basis where $D=$diag$(e^{-i\lambda_{1}},e^{-i\lambda_{2}},e^{-i\lambda_{3}%
},e^{-i\lambda_{4}})$ and $\widetilde{O},O$ are real orthogonal matrices
corresponding to $\widetilde{U}_{A}\otimes\widetilde{V}_{B}$ and $U_{A}\otimes
V_{B}$. Therefore $\mathcal{U}^{T}\mathcal{U}=O^{T}D\widetilde{O}%
^{T}\widetilde{O}DO=O^{T}D^{2}O$. Hence, if we compute the eigenvalues of
$\mathcal{U}^{T}\mathcal{U}$ we will find them to be $\left\{  e^{-2i\lambda
_{1}},e^{-2i\lambda_{2}},e^{-2i\lambda_{3}},e^{-2i\lambda_{4}}\right\}  $.
Taking the arguments of these phases and dividing by two will give us the
$\lambda_{k}$s and via (\ref{transf}) the $\alpha_{k}$s$.$

As an example let us determine the $\alpha_{k}$s for the CNOT gate. In the
computational basis [in the order $\left(  \left|  11\right\rangle ,\left|
10\right\rangle ,\left|  01\right\rangle ,\left|  00\right\rangle \right)  $]
and the magic basis [in the order given by the enumeration in (\ref{mb})] we
find respectively%
\begin{align*}
\mathcal{U}_{CNOT}^{AB}  &  =e^{-i\pi/4}\left(
\begin{array}
[c]{cccc}%
0 & 1 & 0 & 0\\
1 & 0 & 0 & 0\\
0 & 0 & 1 & 0\\
0 & 0 & 0 & 1
\end{array}
\right)  _{CB}\\
&  =\frac{e^{-i\pi/4}}{2}\left(
\begin{array}
[c]{cccc}%
1 & -i & -1 & -i\\
i & 1 & i & -1\\
-1 & -i & 1 & -1\\
i & -1 & i & 1
\end{array}
\right)  _{MB}.
\end{align*}
The overall phase included assures that det$(\mathcal{U}_{CNOT})=1$ and
therefore $\mathcal{U}_{CNOT}\in su(4)$. The eigenvalues of $\mathcal{U}%
_{CNOT}^{T}\mathcal{U}_{CNOT}$ turn out to be $\left\{  i,i,-i,-i\right\}  $.
Taking the square root and then ordering the arguments in decreasing order we
find $\vec{\lambda}=\pi/4\left(  1,1,-1,-1\right)  $. Solving equations
(\ref{transf}) we get $\vec{\alpha}=\pi/4(1,0,0)$.

However in some cases simple algebraic considerations provide a more elegant
way to find the interaction content. We shall demonstrate this on the basis of
the class of controlled-$U$ gates. These gates are of the form $\mathcal{U}%
_{ctrl-U}^{AB}=P_{0}\mathbf{+}P_{1}\mathbf{1}\otimes U$ where $P_{i}=\left|
i\right\rangle _{A}\left\langle i\right|  \otimes\mathbf{1}_{B}$ as we
mentioned in section \ref{basicgates}. If we now take the transpose
$\mathcal{U}_{ctrl-U}^{T}$ in the magic basis and take into account that
$P_{0}^{T}=P_{1}$ and $(\mathbf{1}\otimes U)^{T}=\mathbf{1}\otimes U^{\dagger
}$ we find $\mathcal{U}_{ctrl-U}^{T}\mathcal{U}_{ctrl-U}=\left(
P_{1}\mathbf{+}P_{0}\mathbf{1}\otimes U^{\dagger}\right)  \left(
P_{0}\mathbf{+}P_{1}\mathbf{1}\otimes U\right)  =P_{0}\mathbf{1}\otimes
U^{\dagger}\mathbf{+}P_{1}\mathbf{1}\otimes U=\left|  0\right\rangle
\left\langle 0\right|  \otimes U^{\dagger}\mathbf{+}\left|  1\right\rangle
\left\langle 1\right|  \otimes U$. This operator is block diagonal in the
computational basis and therefore has the same eigenvalues as $U$ but with
multiplicity 2, i.e. has a spectrum $\left\{  e^{i2\beta},e^{i2\beta
},e^{-i2\beta},e^{-i2\beta}\right\}  $ where $e^{\pm i2\beta}$ are the
eigenvalues of $U$. Solving equations (\ref{transf}) for $\lambda_{1}%
=\lambda_{2}=\beta,\lambda_{3}=\lambda_{4}=-\beta$ we find $\vec{\alpha
}=(\beta,0,0)$ $\left[  \vec{\alpha}=(\pi/2-\beta,0,0)\right]  $ for
$\beta\leq\pi/4$ $\left[  \beta\geq\pi/4\right]  $. For the CNOT we have
especially $U=\sigma_{x}$ and thus $\beta=\pi/4$ as it shall be.

\end{document}